\documentclass[prl,twocolumn,showpacs,preprintnumbers,amsmath,amsfonts,amssymb,floatfix,aps,superscriptaddress]{revtex4-2}
\usepackage{graphicx}
\usepackage{xr}
\usepackage{enumitem}
\usepackage{amssymb}
\usepackage{amsmath}
\usepackage{amsfonts}
\usepackage{bm}
\usepackage{dsfont}
\usepackage{comment}
\usepackage{color}
\usepackage{relsize}
\usepackage{float}
\usepackage{bm}
\usepackage[normalem]{ulem}
\usepackage[nodisplayskipstretch]{setspace}
\usepackage[final]{microtype}
\usepackage{tikz}

\usepackage{hyperref}
\hypersetup{
     colorlinks=true,
     linkcolor=blue,
     filecolor=blue,
     citecolor = blue,
     urlcolor=blue,
}
\usepackage[export]{adjustbox}
\newcommand{\op}[1]{\hat{#1}}

\newcommand{\bs}{\boldsymbol}

\newcommand{\be}{\begin{equation}}
\newcommand{\ee}{\end{equation}}

\usepackage{units}

\newcommand{\UIB}{U_{\mathrm{IB}}}
\newcommand{\bk}{\bs{k}}
\newcommand{\br}{\bs{r}}

\newcommand{\GW}{\mathrm{GW}}

\begin{document}
\title{Structure and Dynamics of Bose Polarons across the Mott-Insulator to Superfluid Transition}
\author{Ragheed Alhyder}
\affiliation{Institute of Science and Technology Austria (ISTA), Am Campus 1, 3400 Klosterneuburg, Austria}

\author{Alessio Recati}
\affiliation{Pitaevskii BEC Center, CNR-INO and Dipartimento di Fisica, Universit\`a di Trento, I-38123 Trento, Italy}
\affiliation{Trento Institute for Fundamental Physics and Applications, INFN, 38123 Trento, Italy}

\author{Georg M. Bruun}
\affiliation{Department of Physics and Astronomy, Aarhus University, Ny Munkegade 120, DK-8000 Aarhus C, Denmark}
\begin{abstract}
A mobile impurity particle immersed in a  quantum degenerate  gas leads to the formation of a  quasiparticle, which can 
 serve as a sensor for its environment. 
Here, we develop a unified wave function description of an impurity in a two-dimensional Bose-Hubbard model
across  the Mott-insulator to superfluid transition. Using a multi-mode variational ansatz based on the quantum Gutzwiller approach, we show that 
 the impurity can form several kinds of polaronic and molecular states. In addition to their spectral properties, the 
 wave function provides direct access to the microscopic structure of these states, and we demonstrate 
that strong correlations in the critical regime of the quantum phase transition give rise to several non-analytic 
features in the polaron properties. Increasing the impurity-boson 
interaction leads to transitions between polaron and molecular ground states, and the number of bosons in the impurity dressing cloud grows rapidly at the quantum critical point as the bath modes soften.  
We furthermore reveal qualitatively distinct non-equilibrium many-body dynamics after the injection of the impurity 
in the different phases of the Bose-Hubbard model. Our spectral, real space, and dynamical predictions describe complementary properties of 
  polarons in a strongly correlated bosonic bath, which can be observed using current techniques with atoms in optical lattices. 
\end{abstract}

\maketitle

The experimental and theoretical exploration of highly tunable polarons formed by a mobile impurity atom in a quantum degenerate atomic Fermi gas or Bose-Einstein condensate (BEC) has 
significantly improved our understanding of quasiparticles~\cite{massignanPolaronsAtomicGases2025,grusdtImpuritiesPolaronsBosonic2025,MassReview2014}. Insights and theories 
from these studies have moreover turned out to apply remarkably well to describe polarons formed by excitons interacting with electrons in two-dimensional solid-state 
materials. While the focus has mostly been on impurities in ideal or weakly interacting 
gases, the intriguing topic of strongly correlated environments is gaining increasing attention~\cite{pierce2019a,alhyderImpurityImmersedDouble2020a,alhyderMobileImpurityProbing2022a,vashisht2025,CamachoGuardian2019,Grusdt2016,Heras2020,PhysRevB.110.235302,comaron2025a}.
This is partly due to the possibility of using polarons as sensors for non-trivial quantum phases, which is much needed in 
particular for two-dimensional materials~\cite{massignanPolaronsAtomicGases2025}.  

The Bose-Hubbard model is a canonical case of a strongly correlated system. Here, 
the interplay between on-site repulsion and particle tunneling gives rise to a 
 Mott-insulator to superfluid (MI-SF) transition at integer fillings~\cite{fisher_boson_1989,greiner2002quantum},
 which belongs to the three-dimensional $O(2)$ universality class~\cite{endres_higgs_2012,ranconHiggsAmplitudeMode2014a}. 
 The insulating phase is gapped and incompressible, while the SF supports a BEC with a gapless Goldstone mode and a Higgs mode that 
 becomes gapless at the quantum critical point (QCP). Mobile impurity particles and 
polaron formation have been explored in the lattice boson model~\cite{dingPolaronsBipolaronsTwodimensional2023,santiago-garciaCollectiveExcitationsBose2023,caleffi2021,santiago-garcia2024}, including in the 
quantum critical regime using perturbation theory in the boson-impurity interaction~\cite{colussiLatticePolaronsSuperfluid2023a}, 
which was later generalised to strong interactions using a diagrammatic resummation scheme~\cite{alhyder2025}, 
as well as with Monte Carlo methods~\cite{hartweg2025,cufar2026scaleinvariancepolaronenergy}.

Here, we develop a framework for describing impurity dynamics and polaron formation in the two-dimensional  boson lattice model 
across the full MI-SF transition. It is based on a Quantum Gutzwiller approach~\cite{caleffi2020}
combined with a variational ansatz for the wave function including up to two excitations of the bosons. While 
this approximation is equivalent to the diagrammatic resummation scheme in Ref.~\cite{alhyder2025}, 
having access to the wave function gives complete information regarding the microscopic structure  
of the many-body states both for equilibrium and non-equilibrium dynamics. Using this, we identify 
several intriguing effects on the impurity states in the strongly correlated 
critical region including the emergence of new polaron states, transitions between polaron and molecular ground states, 
and a rapid increase in the number of bosons in the impurity  dressing cloud  at the quantum critical point. 
By solving the time-dependent Schr\"odinger equation, we show that 
the non-equilibrium many-body dynamics ensuing the injection of the impurity is strongly affected by the quantum phase of the surrounding 
bosons. 

\begin{figure*}[t]
\includegraphics[width=\textwidth]{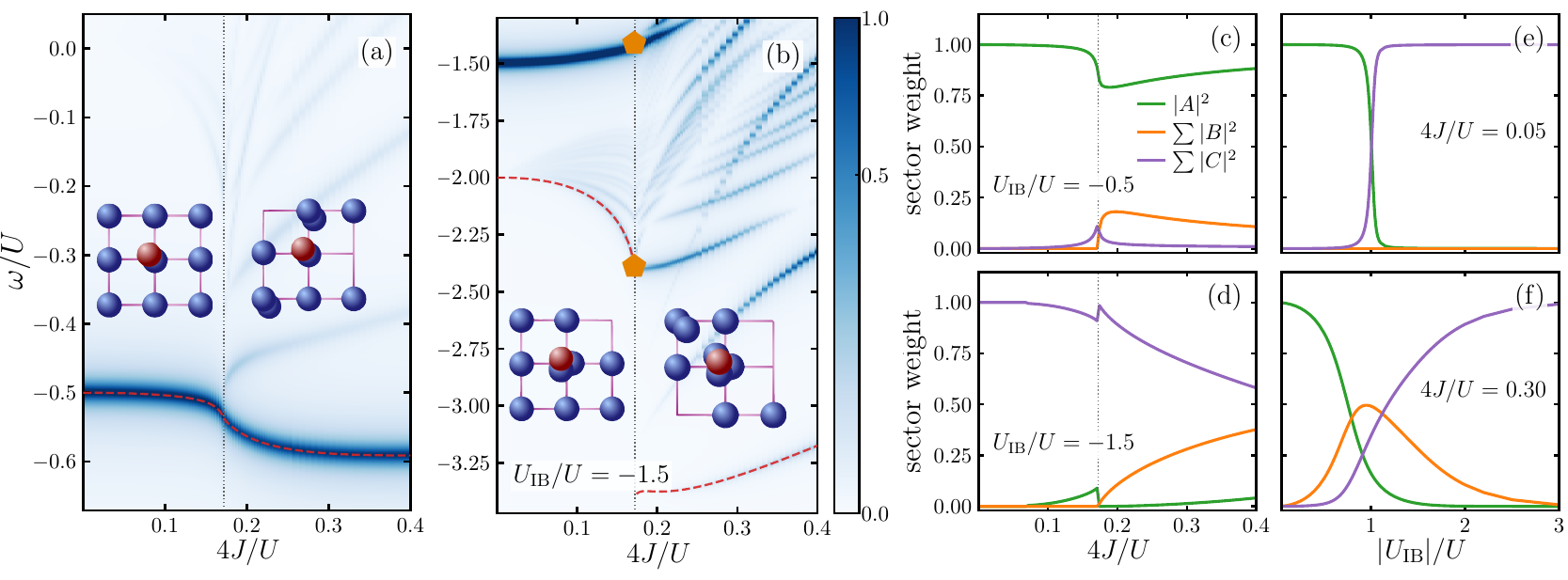}
\caption{\label{fig:fig1}
The impurity spectral function for $U_{\rm IB}/U=-0.5$ (a) and $-1.5$ (b), with the cartoons illustrating the corresponding ground states of the impurity (red ball) 
in a bath of bosons (blue balls). Orange markers indicate the states giving rise to the beating of the Ramsey signal in Fig.~\ref{fig:fig2}, 
 dotted vertical lines give the mean-field critical point $(4J/U)_c=(\sqrt{2}-1)^2\simeq0.172$, and  red dashed lines are the ground state. 
The wave function components are plotted as a function of $4J/U$ for $\UIB/U=-0.5$ (c) and $\UIB/U=-1.5$ (d), and as a function of $|\UIB|/U$  deep in the MI with 
$4J/U=0.05$ (e) and in the SF with $4J/U=0.30$ (f).
} 
\end{figure*}

\textit{Model and variational ansatz.--}We consider a mobile impurity particle interacting with a bath of bosons in a square lattice of $M$ sites described by the Bose-Hubbard model. The Hamiltonian reads 
\begin{align}
\hat{H} &= -J \sum_{\langle \bm{r},\bm{s}\rangle} (\hat{a}^\dagger_{\bm{r}} \hat{a}_{\bm{s}}
+\hat{c}^\dagger_{\bm{r}} \hat{c}_{\bm{s}}+\text{h.c.})- \mu   \sum_{\bm{r}} \hat{n}_{\bm{r}}\nonumber\\
	&+\frac{U}{2}\sum_{\bm{r}} \hat{n}_{\bm{r}}(\hat{n}_{\bm{r}} - 1)  +U_{\text{IB}}\sum_{\bm{r}}\hat{n}_{I,\bm{r}} \hat{n}_{\bm{r}},
	\label{eq:Hamiltonian1}
\end{align}
where  $\hat{a}^\dagger_{\bm{r}}$ ($\hat{c}^\dagger_{\bm{r}}$) are creation  operators for bosons (the impurity) on lattice site $\bm{r}$,  $J$ is the nearest neighbor hopping matrix element taken to be the same for the bosons and the impurity, 
 and $\mu$ is the boson chemical potential. The non-interacting dispersion is $\varepsilon_{\bm{k}} = 4J[\sin^2(k_x/2)+\sin^2(k_y/2)]$, with the band minimum 
 defining zero energy and the lattice constant set to unity.
The second line in Eq.~\eqref{eq:Hamiltonian1} describes the onsite interaction between the bosons with strength 
$U>0$ and between the impurity and the bosons with strength $U_{\mathrm{IB}}$, where $\hat n_{\bf r}=\hat{a}^\dagger_{\bm{r}}\hat{a}_{\bm{r}}$
and $\hat n_{I,\bf r}=\hat{c}^\dagger_{\bm{r}}\hat{c}_{\bm{r}}$. The bath forms either a MI or SF depending 
on the ratio $4J/U$ and $\mu$. 

We treat the bath using the Gutzwiller product state
$
|\mathrm{GW}\rangle
=
\bigotimes_{\mathbf r\in\Lambda}
\sum_{n=0}^{N_{\rm loc}-1}
c_n(\mathbf r)
\frac{(\hat a_{\mathbf r}^{\dagger})^n}{\sqrt{n!}}
|0\rangle_{\mathbf r},
$
with $c_n(\mathbf r)=c_n$ and $\sum_n|c_n|^2=1$ for the translationally invariant ground state considered here. Minimizing the bath energy yields a translationally-invariant ground state with an order parameter $\psi_0 = \langle \hat a_{\bm r}\rangle$ that vanishes in the MI ($c_n = \delta_{n,1}$) and is finite in the SF, recovering the standard Gutzwiller mean-field MI–SF boundary~\cite{KRUTITSKY20161}. Here, we focus on the unit filling case where 
$\bar n=1$.

The QGW method describes fluctuations on top of the ground state by canonical quantization as
$\delta \hat{c}_n({\bf r}) = M^{-1/2} \sum_{\lambda{\bf k}} e^{i \, {\bf k} \cdot {\bf r}} \, (u_{\lambda, {\bf k}, n} \, \hat{b}_{\lambda,{\bf k}} + v_{\lambda, {\bf k}, n} \, \hat{b}^\dag_{\lambda, -{\bf k}})$~\cite{Ripka1985,caleffi2020,alhyder2025}. 
Here, $\hat{b}^\dagger_{\lambda,{\bf k}}$ is a bosonic operator that creates an elementary excitation of the bosons
in the $\lambda^\text{th}$ branch with momentum ${\bf k}$ and energy $\omega_{\lambda,{\bf k}}$.
This allows us to write the bath as a quadratic Hamiltonian $\hat{H}_\mathrm{B} = \sum_{\lambda, {\bf k}} \omega_{\lambda, \mathbf{k}} \, \hat{b}^\dag_{\lambda, \mathbf{k}} \, \hat{b}_{\lambda, \mathbf{k}}$ describing the elementary bath modes throughout the phase diagram. In the SF they include a gapless Goldstone branch and a gapped Higgs branch whose energy softens on approaching the particle-hole-symmetric QCP. In the MI they evolve into gapped particle and hole excitations.

When the impurity-boson interaction is expressed in terms of the $\hat{b}_{\lambda,{\bf k}}$  operators, it leads to terms where the 
impurity scatters on bath excitations as well as terms where it creates/annihilates one or two bath 
excitations~\cite{alhyder2025}, see Sup.\ Mat.~\cite{sm}. 
They  reduce to the standard Bogoliubov couplings deep in the SF and remain well defined throughout the MI-SF transition.
These interactions entangle the impurity with the surrounding bosons, and to describe the resulting many-body state we use 
the variational ansatz 
\begin{align}
|\Psi_{\text{LP}}\rangle  = &\Big(\, A\,\hat c^{\dagger}_{\bs{0}}
+ \sum_{\bk\lambda} B_{\bk}^{\lambda}\,\hat c^{\dagger}_{-\bk}\,\hat b^{\dagger}_{\bk\lambda}\nonumber\\
+& \frac{1}{2}\!\sum_{\bk\lambda,\bk'\lambda'}\! C^{\lambda\lambda'}_{\bk\bk'}\,\hat c^{\dagger}_{-(\bk+\bk')}\,\hat b^{\dagger}_{\bk\lambda}\hat b^{\dagger}_{\bk'\lambda'}\Big)|\mathrm{GW}\rangle.
\label{eq:ansatz}
\end{align}
Here, $\hat c_{\bk}$ is the Fourier transform of $\hat c_{\br}$ and annihilates
 the impurity with lattice momentum $\bk$ and  $A$, $B_{\bk}^{\lambda}$, and $C^{\lambda\lambda'}_{\bk\bk'} = C^{\lambda'\lambda}_{\bk'\bk}$ are coefficients describing 
 the impurity together with zero, one, or two bath excitations. We focus on the ground state with 
 vanishing total momentum. The wave function $|\Psi_{\text{LP}}\rangle$ is a Chevy-type variational 
 ansatz~\cite{Chevy2006} generalised to include strong correlations between the bosons via the Gutzwiller ansatz. It also includes up to two 
 bath excitations as this is necessary to describe the binding of a boson to the impurity for strong interactions in the MI regime as we shall see below. The wave function interpolates between the impurity being dressed with  Bogoliubov modes in the SF and with particle-hole 
 excitations in the MI. Truncating to two bosonic excitations in Eq.~\eqref{eq:ansatz}  is 
  equivalent to the diagrammatic approach in Ref.~\cite{alhyder2025}, but here we have access to the wave functions as well giving 
  complete information regarding the many-body dynamics.


\textit{Spectrum and wave functions.--}We now analyze the many-body spectrum and corresponding eigenstates obtained by 
diagonalising the Hamiltonian projected onto the variational subspace of Eq.~\eqref{eq:ansatz}. 
Numerical details and convergence checks are given in the Sup.\ Mat.~\cite{sm}.
From the energies $E_n$ and normalized eigenstates $|\psi_n\rangle$, we can calculate 
the impurity spectral function as 
$
\mathcal A(\omega,4J/U)
=
2\pi\sum_n
\left|\langle\Psi(0)|\psi_n\rangle\right|^2
\delta(\omega-E_n),
$
where $|\Psi(0)\rangle=\hat c^\dagger_{\bs 0}|\mathrm{GW}\rangle$ and $\langle\Psi(0)|\psi_n\rangle=A_n$ from
 Eq.~\eqref{eq:ansatz} so that each state has residue $|A_n|^2$ with $\sum_n|A_n|^2=1$. 

Figure~\ref{fig:fig1}(a)-(b) show the spectral function 
as a function of the ratio $4J/U$ of the boson hopping and repulsion for impurity-boson interaction $\UIB/U=-0.5$ and  $\UIB/U=-1.5$. 
 Panel (a) shows that for weak impurity-boson interaction $\UIB/U=-0.5$,
the spectral function has a single well-defined polaron branch with the mean-field energy $\UIB\bar n=\UIB$ corresponding to a bare impurity 
moving in an essentially unchanged incompressible 
MI phase.  We see that the energy abruptly decreases at the QCP reflecting that the bath develops gapless modes and becomes much more compressible. 
The energy continuously decreases 
in the highly compressible SF phase where the ground state evolves into a usual Bose polaron consisting of the impurity dressed by Bogoliubov modes~\cite{dingPolaronsBipolaronsTwodimensional2023}.

To analyse this  further, we plot in Fig.~\ref{fig:fig1}(c) the wave function components of the polaron ground state. 
The residue $|A|^2$ is close to unity in the MI and decreases sharply at the QCP, 
explicitly showing that the dressing of the impurity increases with the compressibility of the bath.

\begin{figure}[t]
\centering
\includegraphics[width=1.0\columnwidth]{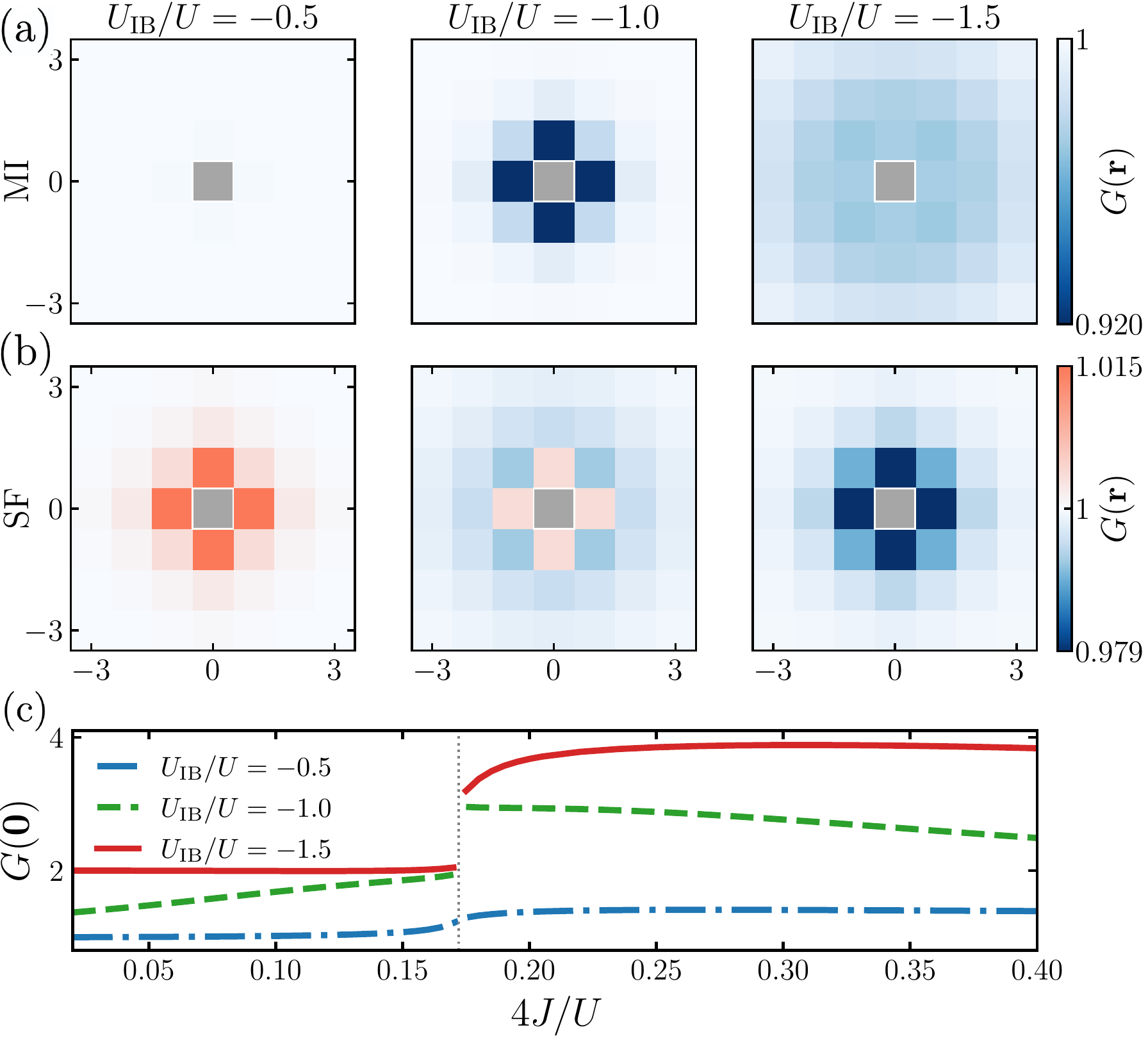}
\caption{\label{fig:fig3}
Boson density around the impurity given by $G(\br)$ for $\UIB/U=-0.5$, $-1.0$, and $-1.5$ 
deep in the MI with $4J/U=0.05$ (a), and in the SF  with $4J/U=0.30$ (b). The gray square marks the impurity site, whose density is shown separately in panel (c). Blue (red) indicates a boson density below (above) the bulk value $\bar n=1$.
(c) On-site impurity-conditioned density $G(\bs 0)$ of the ground  state as a function of $4J/U$ for $\UIB/U=-0.5$ in blue, $-1.0$ in green, and $-1.5$ in red.
}
\end{figure}

For the case of strong impurity-boson interaction $\UIB/U=-1.5$ shown in Fig.~\ref{fig:fig1}(b), the ground state energy (dashed line) is $-2U$
 deep in the MI phase. This comes from the  impurity binding an extra boson so that its energy is approximately
$2\UIB+U= -2U$. Consistent with this, panel (d) shows that the wave function is dominated by the $C$
terms describing the particle-hole excitations of the MI forming an extra boson at the impurity site. This molecular (dimer) ground state has 
a small residue making it only faintly visible in the spectrum. Instead, a polaron consisting essentially of a bare impurity with the 
mean-field energy $\UIB\bar n= -1.5U$ takes most of the spectral weight. 
 Approaching the quantum phase transition from the MI, strong coupling to the Goldstone and the Higgs modes that become gapless at the QCP
decreases the ground state  energy. At the QCP, the lowest state switches to a more strongly dressed branch with energy $\simeq-3.4U$, accompanied by a sharp redistribution of the zero-, one-, and two-excitation weights.
This leads to discontinuities in the corresponding ground state wave function components shown in panel (d) with the 
single excitation weight $\sum|B|^2$  continuously growing in the  SF.

To further explore the microscopic structure of the many-body states, we plot in Fig.~\ref{fig:fig1}(e)-(f)
the ground state wave function components  as a function of the impurity-boson interaction strength $U_\text{IB}<0$.  
Deep in the MI phase with $4J/U=0.05$ [panel (e)], the ground state undergoes a sharp transition 
 around $|\UIB|/U=1$ from an essentially bare impurity state with $|A|^2 \simeq 1$ to a dimer state where the impurity has bound one 
 extra boson to its lattice site so that $\sum|C|^2 \simeq 1$. This can straightforwardly be understood from the 
atomic MI limit where particle hopping vanishes $J=0$.  Indeed, the energy of a site containing the impurity and $m$ bosons is 
 $E_m=Um(m-1)/2+m\UIB$ so that it becomes energetically favorable to bind one more boson when $E_{m+1}-E_m=mU+\UIB\le 0$. 
 It follows that the number of bosons at the impurity site increases from $m$ to $m+1$ when $|\UIB|$ crosses $mU$. 
 In the SF phase shown in panel (f),  the  transition of the polaron wave function is  smooth with a maximum
 in the single excitation component 
  $\sum|B|^2$  near $|\UIB|/U\sim 1$. 

\textit{Polaron cloud.--}Motivated by the ability to detect atoms with single site resolution in optical lattices using quantum gas microscopy, we now explore the real space impurity-density correlation function~\cite{penaArdilaImpurity2015,grusdtImpuritiesPolaronsBosonic2025}
\begin{equation}
G(\br) \;=\; \frac{\langle\Psi_{\rm LP}|\,\hat n_I(\bs 0)\,\hat n(\br)\,|\Psi_{\rm LP}\rangle}{\langle\Psi_{\rm LP}|\,\hat n_I(\bs 0)\,|\Psi_{\rm LP}\rangle}.
\label{eq:cloud-def}
\end{equation}
This   gives the boson density at lattice site $\br$ conditioned on finding the impurity at $\bs 0$ and is  directly available from the
wave function $|\Psi_{\text{LP}}\rangle$~\cite{sm}. For the translationally invariant state considered here, $\langle\hat n_I(\bs 0)\rangle=1/M$ and $G(\br)$ approaches the filling $\bar n=1$ far from the impurity. The correlation function $G(\br)$ can be measured with a quantum-gas microscope by post-selecting configurations depending on the impurity site~\cite{Bakr,Sherson2010,koepsellImagingMagneticPolarons2019}. 

In Fig.~\ref{fig:fig3} (a) and (b), we plot $G(\br)$ for two different impurity-boson interaction strengths when the bosons are 
deep in the MI phase with $4J/U=0.05$, and in the SF phase with 
$4J/U=0.3$. The impurity site is masked in the spatial maps and shown separately in panel (c) due to its large density.
In the MI phase in panel (a), the charge gap suppresses the density response and the local cloud is governed by the discrete on-site occupations of the impurity site. At $\UIB/U=-0.5$ the impurity sits inertly on a Mott site and $G(\bs 0) \simeq 1$, while at $\UIB/U=-1.5$ the impurity binds an extra boson giving $G(\bs 0) \simeq 2$
consistent with our discussion above regarding the spectral function. We also see that the extra boson on the impurity site is compensated by a shallow depletion of the surrounding sites.
At \(U_{\rm IB}/U=-1.0\), the one- and two-boson impurity-site configurations are degenerate, so finite tunneling resonantly transfers a neighboring boson onto the impurity, resulting in an on-site enhancement and a depletion of the neighboring sites, forming a localized particle-hole redistribution. The depletion of the neighboring sites is largest at this interaction strength~\cite{sm}.

In panel (b), it is shown that in the SF, the large  compressibility permits a smooth density response
with $G(\bs 0)$ gradually increasing with the impurity-boson interaction strength from  $1.4$ to $3.9$ 
corresponding to nearly three extra 
bosons on the impurity site on average. 
As seen in Fig.~\ref{fig:fig1}(f), this transition is a result of the impurity mainly being dressed by a single 
excitation for small $|\UIB|$ interactions whereas it is mainly dressed by two excitations for large 
$|\UIB|$. Close to the impurity, the density of the neighboring sites is enhanced for weak interactions, as expected for an attractive impurity in a compressible bath~\cite{penaArdilaImpurity2015}, whereas it is depleted for large $|\UIB|/U$ where the bosons accumulated on the impurity site are taken from its immediate surroundings.
In the Sup.~Mat. it is  explicitly shown how $G(\br)$ is determined by the bath compressibility in the perturbative regime~\cite{sm}.
We also plot in the Sup.\ Mat.\  an  average $\overline{G}(r) = N(r)^{-1}\sum_{|\br'|\le r} G(\br')$ of the 
impurity dressing cloud where $N(r)$ is the number of lattice sites within radius $r$, as well as the density $G(\mathbf e_x)$ on the sites neighboring the impurity as a function of $\UIB$.

In Fig.~\ref{fig:fig3} (c), we plot the boson density $G(\mathbf {0})$ on the impurity 
site as a function of $4J/U$ for the same three impurity-boson interaction strengths as in Fig.~\ref{fig:fig3}. The on-site response evolves smoothly for \(U_{\rm IB}/U=-0.5\), develops a resonance-driven maximum near the transition at \(U_{\rm IB}/U=-1\), and switches to a more strongly dressed branch for \(U_{\rm IB}/U=-1.5\). 
The soft critical modes at the QCP facilitate these rearrangements, while their coupling dependence is controlled by the polaron branch structure. We also see that there is a jump 
in the boson density at the impurity site at the QCP due to these gapless modes.

\begin{figure}[t]
\centering
\includegraphics[width=\columnwidth]{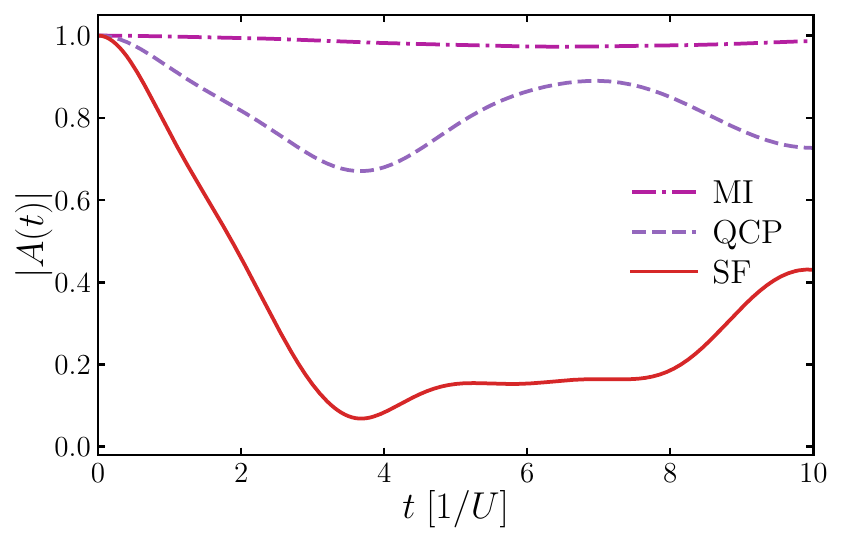}
\caption{\label{fig:fig2}
Ramsey contrast $|A(t)|$ following the injection of a zero momentum impurity at time $t=0$ for 
 $U_{\rm IB}/U=-1.5$  in the MI with $4J/U=0.05$ (dash-dotted), near the QCP  with $4J/U=0.175$ (dashed), and in the SF with 
 $4J/U=0.30$ (solid).
}
\end{figure}

\textit{Polaron formation dynamics.}-- The non-equilibrium  
dynamics of  polaron formation after a sudden injection of an impurity into a continuum BEC has been 
investigated using Ramsey spectroscopy\cite{skouNonequilibriumQuantumDynamics2021,etrych2024universal,alhyder2026a}. Such experiments probe the 
time-dependence of the impurity Ramsey contrast $|\langle\Psi(0)|\Psi_{\text{LP}}(t)\rangle| = |A(t)|$, 
 with the initial value $A(0)=1$ corresponding to the injection of the impurity at time $t=0$. 
Within the variational ansatz, the time-evolved state $|\Psi_{\text{LP}}(t)\rangle = e^{-i\hat H t}|\Psi(0)\rangle$ is obtained by an Arnoldi/Lanczos time-stepper~\cite{sm}, so that $|A(t)|$ is a finite sum over the eigen-frequencies of $\tilde H$ weighted by the bare-impurity overlap of each eigenstate. 
Figure~\ref{fig:fig2} plots $|A(t)|$ for $\UIB/U=-1.5$ for $4J/U=0.05$, $0.175$, and $0.30$.

In the MI ($4J/U=0.05$), the dynamics is straightforward: the host charge gap protects the bare impurity from low-energy dressing so that the  residue stays close to one. 
This corresponds to the mean-field polaron  with large 
spectral weight and energy $ \UIB\bar n=-1.5U$ shown in Fig.~\ref{fig:fig1}(b). The residual modulation of $|A(t)|$
comes from a small admixture of the impurity with the dimer ground state with 
 energy close to $-2U$, see Fig.~\ref{fig:fig1}(b).

The dynamics changes qualitatively near the QCP, where the contrast develops a clear oscillation with a first minimum near $tU\simeq3.7$ and a revival near $tU\simeq7$. Since $A(t)=\sum_n|A_n|^2 e^{-iE_n t}$, this is the beat between the bright mean-field polaron with energy near $-1.4U$ and a large overlap with the bare impurity state 
$\hat c^\dagger_{\mathbf 0}|\text{GW}\rangle$ created at $t=0$,
and the attractive branch near $-2.3U$. These states are indicated with orange markers in Fig.~\ref{fig:fig1}(b), and their splitting $\Delta E\simeq 0.9 \,U$ sets the minimum at $\pi/\Delta E$ and the revival at $2\pi/\Delta E$. The revival is incomplete because the spectral weight of the attractive branch is spread over several nearby states.

In the SF,  the contrast quickly decreases to $|A(t)|\simeq 0.1$ before partially recovering. This is the dynamical hallmark of strong many-body dressing in a compressible host~\cite{nielsen2019,skouNonequilibriumQuantumDynamics2021}.
The impurity excites a continuum of low-energy excitations in the bath leading to decoherence as it 
dynamically attracts bosons forming the ground state~\cite{dingPolaronsBipolaronsTwodimensional2023}. 
The partial revival of $|A(t)|$ in Fig.~\ref{fig:fig2} reflects residual beating between different eigenstates. 
The sequence in Fig.~\ref{fig:fig2} therefore tracks a smooth progression from an essentially bare (mean-field) impurity in the MI, through oscillations between different states in the critical regime, to the formation of a strongly dressed ground state polaron in the SF.

\textit{Conclusion and outlook.--}
By combining a  quantum Gutzwiller approach  with a  variational ansatz,
we developed a unified wave function for a mobile impurity in a two-dimensional lattice Bose gas across  the MI-SF quantum phase transition.
This describes the entanglement of the impurity with bosonic modes leading to 
 the formation of several kinds of polarons and molecular states. 
We showed that strong correlations in  the critical region 
of the MI-SF transition lead to several interesting effects that are directly available from the 
wave function. This includes polaron to dimer transitions, the emergence of new states  as well as  a strong rearrangement of the local density around the impurity near the QCP.
We furthermore identified distinct dynamical regimes in the non-equilibrium dynamics ensuing the injection of an impurity. 
All these results concern complementary spectral, real space, and dynamical properties of Bose polarons, which can be observed experimentally with atoms 
in optical lattices.

Several interesting research directions follow naturally. This  includes exploring the induced interaction between two impurities and the possible formation of bipolaron states across the MI--SF transition~\cite{camacho-guardianBipolaronsBoseEinsteinCondensate2018a,casteelsBipolaronsMultipolaronsConsisting2013,Camacho-GuardianLandauEffectiveInteraction2018}. 
 Moreover, adding nearest-neighbor interactions would generalise the present approach to the extended Bose-Hubbard model, enabling studies of impurity dressing across charge-density-wave and supersolid phases~\cite{Rossini2012,Sengupta2005}. On the experimental side, 
 it would be very interesting to explore the  predicted 
 properties of Bose polarons using radio-frequency or Ramsey spectroscopy, and quantum gas microscopy
  with ultracold atoms.

 \paragraph*{Acknowledgments.}

This work has been supported by the Provincia autonoma di Trento, the INFN through the RELAQS project, the Italian Ministry of University and Research (MUR) through the PNRR MUR project: `National Quantum Science and Technology Institute' - NQSTI (PE0000023).
G.M.B.\ acknowledges support from the Novo Nordisk Foundation (grant no.
NNF23OC0086599). R. A. received funding from the Austrian Academy of Sciences ÖAW grant No. PR1029OEAW03. 

\bibliographystyle{apsrev4-2}
\bibliography{references}


\clearpage
\onecolumngrid
\begin{center}
{\large\textbf{Supplemental Material:
Structure and Dynamics of Bose Polarons across the Mott-Insulator to Superfluid Transition}}\\[8pt]
Ragheed Alhyder, Alessio Recati, Georg M. Bruun
\end{center}
\vspace{6pt}
\twocolumngrid

\setcounter{section}{0}
\setcounter{equation}{0}
\setcounter{figure}{0}
\renewcommand{\thesection}{S\Roman{section}}
\renewcommand{\theequation}{S\arabic{equation}}
\renewcommand{\thefigure}{S\arabic{figure}}

\renewcommand{\theHsection}{S\Roman{section}}
\renewcommand{\theHequation}{S\arabic{equation}}
\renewcommand{\theHfigure}{S\arabic{figure}}
\renewcommand{\theHtable}{S\arabic{table}}

\section{Gutzwiller bath}
\label{sec:bath}

We treat the host described by Eq.~(\ref{eq:Hamiltonian1}) of the main text within the Gutzwiller (GW) approximation. The bath state is a translationally invariant on-site product
\begin{equation}
|\GW\rangle \;=\; \prod_{i=1}^{M}\,\sum_{n=0}^{N_{\rm loc}-1} c_n\,|n\rangle_i,
\quad \sum_{n=0}^{N_{\rm loc}-1}|c_n|^2 = 1,
\label{eq:GWvacuum}
\end{equation}
with $M = L_x L_y$ lattice sites and a local on-site cutoff $N_{\rm loc}$. Throughout the Supplemental Material we set $U=1$ and measure all energies, including $\mu$, in units of $U$. Here $J$ is the nearest-neighbor hopping of Eq.~(\ref{eq:Hamiltonian1}) of the main text, so that $4J$ is the mean-field hopping on the square lattice. The amplitudes $\{c_n\}$ are obtained by minimizing the bath energy at fixed chemical potential. This reduces to a self-consistent eigenvalue problem for the on-site $N_{\rm loc}\!\times\!N_{\rm loc}$ matrix
\begin{align}
H^{(\GW)}_{nm}(\psi_0) \;=\;& \Big[\tfrac{1}{2} n(n-1) - \mu\, n\Big]\delta_{nm} \nonumber\\
& -\, 4J\,\psi_0 \big(\sqrt{n}\,\delta_{n,m+1} + \sqrt{n+1}\,\delta_{n+1,m}\big),
\label{eq:Hgw}
\raisetag{35pt}
\end{align}
parameterized by the order parameter $\psi_0 = \langle\GW|\op b_i|\GW\rangle = \sum_n \sqrt{n}\,c_{n-1}^* c_n$, which is determined self-consistently. The MI corresponds to $c_n = \delta_{n,1}$ (vanishing $\psi_0$) and the SF to a finite $\psi_0$ with the on-site distribution broadened across several $n$. We denote the GW density by $\bar n = \sum_n n\,|c_n|^2$ and work along the line of unit filling, adjusting $\mu$ at each $4J/U$ such that the density including the fluctuation correction, $\bar n_{\rm eff}$ of Eq.~\eqref{eq:nbar_eff}, is equal to one within $10^{-3}$.

We define site-local fluctuation operators $\delta\op c_n^{(i)} = \op c_n^{(i)} - c_n$ and impose the Gauss-like constraint $c_n^* \delta\op c_n + \mathrm{h.c.} = 0$ that keeps the local norm. The quadratic Hamiltonian governing the fluctuations is diagonal in momentum because $|\GW\rangle$ is translationally invariant. For each lattice momentum $\bk$ in the Brillouin zone, the linearized dynamics defines a $2 N_{\rm loc} \!\times\! 2 N_{\rm loc}$ Bogoliubov-de-Gennes (BdG) problem
\begin{equation}
\begin{pmatrix} A(\bk) & B(\bk) \\ -B^{*}(\bk) & -A^{*}(\bk) \end{pmatrix}\!
\begin{pmatrix} u_{\bk\lambda} \\ v_{\bk\lambda} \end{pmatrix}
\;=\; \omega_{\bk\lambda}\!
\begin{pmatrix} u_{\bk\lambda} \\ v_{\bk\lambda} \end{pmatrix},
\label{eq:BdG}
\end{equation}
with the matrix blocks
\begin{align}
A_{nm}(\bk) &= \big[\tfrac{1}{2}\,n(n-1) - \mu\,n - \omega_0\big]\delta_{nm} \nonumber\\
&\quad - 4J\,\big[\psi_0^*\sqrt{n}\,\delta_{n,m+1} + \psi_0\sqrt{m}\,\delta_{n+1,m}\big] \nonumber\\
&\quad - J_{\bk}\,\big[\sqrt{n m}\, c_{m-1} c_{n-1} \nonumber\\
&\qquad\quad + \sqrt{(n+1)(m+1)}\, c_{m+1} c_{n+1}\big], \label{eq:Ablock}
\end{align}
\begin{align}
B_{nm}(\bk) &= - J_{\bk}\,\big[\sqrt{n(m+1)}\, c_{m+1} c_{n-1} \nonumber\\
&\qquad\quad + \sqrt{(n+1)\,m}\, c_{m-1} c_{n+1}\big], \label{eq:Bblock}
\end{align}
where $\omega_0 = \langle\GW|\op H_{\mathrm{bath}}|\GW\rangle$ is the per-site GW energy and the lattice form factor is
\begin{align}
J_{\bk} \;&=\; 4J\,\big[\,1 - \sin^2(k_x/2) - \sin^2(k_y/2)\,\big] \nonumber\\
&=\; 2J\,(\cos k_x + \cos k_y).
\end{align}
The eigenvectors $(u_{\bk\lambda},v_{\bk\lambda})$ define Bogoliubov phonons $\op b^\dagger_{\bk\lambda}$ labelled by momentum $\bk$ and band index $\lambda = 0,1,\dots,N_{\rm loc}-1$. They are orthonormal in the symplectic norm
\begin{equation}
\sum_{n}\big(\,u^*_{n,\bk\lambda} u_{n,\bk\lambda'} - v^*_{n,\bk\lambda} v_{n,\bk\lambda'}\,\big) = \delta_{\lambda\lambda'}.
\end{equation}
The index $\lambda=0$ labels a spurious gauge mode enforced by the local normalization constraint; it carries no physical weight and is excluded from the dynamics. Among the physical branches $\lambda=1,\dots,N_{\rm loc}-1$, the lowest reproduces the gapless Goldstone mode in the SF (gapped by the Mott gap in the MI) and the next carries the gapped amplitude (Higgs) mode that softens on approaching the QCP; the remaining branches lie at order $U$. We keep the lowest $\lambda_{\rm cut}-1$ physical branches, so that $\lambda_{\rm cut}=6$ retains five branches per momentum and, at $L=10$, $N_{\text{mode}} = M(\lambda_{\rm cut}-1)=500$.

\section{Variational ansatz and impurity-bath vertices}
\label{sec:vertices}

The variational state (Eq.~(\ref{eq:ansatz}) of the main text) reads, in the reduced label notation $s \equiv (\bk,\lambda)$ and at total polaron momentum $\bs{P}=\bs{0}$,
\begin{align}
|\psi\rangle =& \Big[\,A\,\op c^\dagger_{\bs 0}
+ \sum_{s} B_s\, \op c^\dagger_{-\bk_s}\,\op b^\dagger_{s}\nonumber\\
&+ \tfrac{1}{2}\!\sum_{s,t} C_{s,t}\,\op c^\dagger_{-(\bk_s+\bk_t)}\,\op b^\dagger_{s}\op b^\dagger_{t}\,\Big]|\GW\rangle,
\label{eq:ansatzS}
\end{align}
with the symmetry constraint $C_{s,t}=C_{t,s}$ and the impurity creation operator $\op c^{\dagger}_{\bk}$ at lattice momentum $\bk$. The impurity momentum is fixed in every term by overall momentum conservation. Higher excitation sectors can encode qualitatively important correlations rather than merely improve quantitative accuracy. A related example occurs for continuum Fermi polarons, where higher-order processes generate logarithmic beyond-mean-field corrections to the energy~\cite{alhyder2024}.

Projecting the impurity-bath interaction $\UIB \sum_i \op n_i \op n^I_i$ onto the variational subspace generates four vertices, all built from the BdG amplitudes $(u_{n,\bk\lambda},v_{n,\bk\lambda})$ and the local densities $\delta n_n \equiv n - \bar n$,
\begin{align}
N_{s} &= \sum_n n\,\big(\,c_n^{*}\,u_{n,s} + c_n\,v_{n,s}\,\big),
\label{eq:Nvertex}\\
U_{s,t} &= \sum_n (n-\bar n)\,u^{*}_{n,s} u_{n,t},
\label{eq:Uvertex}\\
V_{s,t} &= \sum_n (n-\bar n)\,v^{*}_{n,s} v_{n,t},
\label{eq:Vvertex}\\
W_{s,t} &= \sum_n (n-\bar n)\,u^{*}_{n,s} v_{n,t}.
\label{eq:Wvertex}
\end{align}
Physically, $N_s$ is the linear matrix element coupling the impurity to a single bath phonon, $U_{s,t}$ and $V_{s,t}$ are the normal one-body kernels (particle-particle and hole-hole), and $W_{s,t}$ is the anomalous (pair-creation) kernel. We use throughout the symmetric pair vertex $W^{\rm sym}_{s,t} = W_{s,t}+W_{t,s}$ and the combined kernel $K_{s,t} = U_{s,t} + V_{t,s}$, which is the natural object appearing in normal one-body scattering on top of a Bogoliubov vacuum.

The time-independent variational equations of motion, projected on the components of Eq.~\eqref{eq:ansatzS}, take the form
\begin{align}
\omega A \;=\;& \UIB\,\bar n_{\rm eff}\,A
+ \frac{\UIB}{M}\sum_s N_s^{*}\,B_s \nonumber\\
&+ \frac{\UIB}{2M^{2}}\sum_{s,t} W^{\rm sym\,*}_{s,t}\,C_{s,t},
\label{eq:eomA}\\
\frac{\omega}{M}\, B_s \;=\;& \frac{1}{M}\,(\,\varepsilon_{\bk_s} + \UIB\bar n + \omega_s\,)\,B_s
+ \frac{\UIB}{M}\,N_s\,A \nonumber\\
&\!+ \frac{\UIB}{M^{2}}\sum_t K_{s,t}\,B_t
+ \frac{\UIB}{2M^{2}}\sum_t N_t^{*}\,(C_{s,t}+C_{t,s}),
\label{eq:eomB}
\end{align}
\begin{align}
\frac{\omega}{2M^{2}}\, C_{s,t} \;=\;& \frac{1}{2 M^{2}}\,(\,\varepsilon_{\bk_s+\bk_t} + \UIB\bar n + \omega_s + \omega_t\,)\,C_{s,t} \nonumber\\
&\!+ \frac{\UIB}{2 M^{2}}\,W^{\rm sym}_{s,t}\,A + \frac{\UIB}{2 M^{2}}\,(\,N_s B_t + N_t B_s\,) \nonumber\\
&\!+ \frac{\UIB}{4 M^{3}}\,\big[\,(K\!\cdot\!S)_{s,t} + (K\!\cdot\!S)_{t,s}\,\big],
\label{eq:eomC}
\end{align}
where $S_{s,t} = C_{s,t}+C_{t,s}$, $\varepsilon_{\bk} = 4J\,[\sin^2(k_x/2)+\sin^2(k_y/2)]$ is the impurity dispersion, and the renormalized background is
\begin{equation}
\bar n_{\rm eff} \;=\; \bar n + \frac{1}{M}\sum_s V_{s,s}.
\label{eq:nbar_eff}
\end{equation}
The factor of $1/M$ in front of every internal momentum sum is the discrete-grid form of the continuum $\int d^2k/(2\pi)^2$ that one would have on an infinite lattice.

\section{Generalized eigenvalue problem and Krylov solution}
\label{sec:metric}

The basis vectors $\op c^\dagger_{\bs 0}|\GW\rangle$, $\op c^\dagger_{-\bk_s}\op b^\dagger_s|\GW\rangle$, and $\op c^\dagger_{-(\bk_s+\bk_t)}\op b^\dagger_s\op b^\dagger_t|\GW\rangle$ are mutually orthogonal, but the variational amplitudes $A$, $B_s$, $C_{s,t}$ carry sector-dependent normalization weights, namely a lattice plane-wave factor $1/M$ per phonon, and a Chevy $1/2$ from storing $C_{s,t}$ on the full $N_s\!\times\!N_s$ grid (where each physical pair $\op b^\dagger_s\op b^\dagger_t = \op b^\dagger_t\op b^\dagger_s$ is counted twice). Within the variational principle this turns the stationarity condition into the generalized eigenvalue problem
\begin{equation}
\op H\,|\psi\rangle \;=\; \omega\,\op S\,|\psi\rangle,
\label{eq:Sgen}
\end{equation}
where $\op H$ is the Hamiltonian projected onto the ansatz and $\op S$ is the diagonal metric
\begin{equation}
\op S \;=\; \mathrm{diag}\!\left(\,1\,;\,\tfrac{1}{M}\,\delta_{s s'}\,;\,\tfrac{1}{2 M^2}\,\delta_{s s'}\delta_{t t'}\,\right)
\label{eq:Sdiag}
\end{equation}
in the sector decomposition $(A,B_s,C_{s,t})$. Because $\op S>0$, we form the symmetric square root
\begin{equation}
T \;\equiv\; \op S^{1/2},\qquad
\widetilde{H} \;\equiv\; T^{-1}\,\op H\,T^{-1},
\label{eq:Htilde}
\end{equation}
and recover an ordinary Hermitian eigenvalue problem
\begin{equation}
\widetilde{H}\,|\widetilde\psi\rangle \;=\; \omega\,|\widetilde\psi\rangle,\qquad |\psi\rangle = T^{-1}|\widetilde\psi\rangle.
\label{eq:Hstd}
\end{equation}
The components of $|\widetilde\psi\rangle$, $\widetilde A = A$, $\widetilde B_s = B_s/\sqrt{M}$ and $\widetilde C_{s,t} = C_{s,t}/(\sqrt{2}M)$, satisfy $|\widetilde A|^2+\sum_s|\widetilde B_s|^2+\sum_{s,t}|\widetilde C_{s,t}|^2=1$ and define the sector weights shown in Fig.~\ref{fig:fig1} of the main text.
$\widetilde H$ is finite-dimensional, with dimension $D = 1 + N_{\text{mode}} + N_{\text{mode}}^{2}$, but $D$ grows quickly with the lattice size. At $L=10$, $\lambda_{\rm cut}=6$, we have $N_s = 500$ and $D \simeq 2.5\times 10^{5}$. Forming $\widetilde H$ explicitly is therefore impractical, and we instead represent it through a matrix-vector product
\begin{equation}
\widetilde H \mathbf{x} \;=\; T^{-1}\,\op H\,(\,T^{-1}\,\mathbf{x}\,),
\label{eq:Htilde_mv}
\end{equation}
in which $\op H$ acts via the closed-form Eqs.~\eqref{eq:eomA}--\eqref{eq:eomC} and $T^{-1}$ is a trivial diagonal multiplication thanks to Eq.~\eqref{eq:Sdiag}.

The lowest few eigenpairs of $\widetilde H$ are then obtained by a Krylov-subspace iteration. We use the Implicitly Restarted Arnoldi Method, which specializes to the Implicitly Restarted Lanczos Method since $\widetilde H$ is Hermitian, with a starting vector $\mathbf{v}_0 = \op c^{\dagger}_{\bs 0}|\GW\rangle$ (i.e.~the bare-impurity component) so that the ground state is locked into the connected polaron sector. Convergence of the ground state to relative tolerance $10^{-6}$ requires fewer than $100$ matrix-vector products in all parameter regimes we consider, including the QCP, and the lowest $100$ eigenpairs require at most a few thousand. The variational spectral function $\mathcal{A}$ shown in Fig.~\ref{fig:fig1} of the main text is constructed by extracting the lowest $100$ eigenvalues and broadening each by a Lorentzian of width $\eta = 0.01\,U$, with weights given by the squared overlap $|\langle\GW|\op c_{\bs 0}|\widetilde\psi_n\rangle|^2$ between each eigenstate and the bare impurity.

\section{Time evolution and Ramsey contrast}
\label{sec:dynamics}

The Ramsey contrast $\mathcal C(t) = |\langle\GW|\op c_{\bs 0}|\psi(t)\rangle|$ of the main text follows from the same operator $\widetilde H$ via a unitary time evolution. Starting from the bare-impurity state $|\psi(0)\rangle = \op c^{\dagger}_{\bs 0}|\GW\rangle$ -- i.e.~the variational vector $(A=1,\,B_s=0,\,C_{s,t}=0)$ -- and applying the metric-corrected evolution
\begin{equation}
i\,\partial_t (T|\psi(t)\rangle) \;=\; \widetilde H\,(T|\psi(t)\rangle),
\label{eq:tevol}
\end{equation}
the bare-impurity amplitude evolves as
\begin{equation}
A(t) \;=\; \mathbf e_A^{\dagger}\,e^{-i\widetilde H t}\,\mathbf e_A,
\label{eq:Aoft}
\end{equation}
where $\mathbf e_A$ is the unit vector along the bare-impurity ($A$-sector) component in the transformed basis; since the metric is trivial in that sector, $T\,\op c^{\dagger}_{\bs 0}|\GW\rangle = \mathbf e_A$.

We integrate Eq.~\eqref{eq:tevol} by a time-stepped Arnoldi/Lanczos scheme. At each step of size $\Delta t$ we build the Krylov subspace
\begin{equation}
\mathcal K_m \;=\; \mathrm{span}\!\big\{\,|\widetilde\psi\rangle,\,\widetilde H|\widetilde\psi\rangle,\,\widetilde H^{2}|\widetilde\psi\rangle,\,\dots,\,\widetilde H^{m-1}|\widetilde\psi\rangle\,\big\}
\end{equation}
of dimension $m \ll D$. Since $\widetilde H$ is Hermitian, the Arnoldi recursion reduces to a Lanczos recursion, and we run it with full reorthogonalization to produce the orthonormal basis $\{\,|\op q_k\rangle\,\}_{k=0}^{m-1}$ and the tridiagonal projection
\begin{equation}
T_m \;\equiv\; Q_m^{\dagger}\,\widetilde H\,Q_m, \qquad Q_m = (|\op q_0\rangle,\dots,|\op q_{m-1}\rangle).
\end{equation}
The propagator over the step $\Delta t$ is then approximated by
\begin{equation}
|\widetilde\psi(t+\Delta t)\rangle \;\approx\; Q_m\,e^{-i T_m \Delta t}\,Q_m^{\dagger}|\widetilde\psi(t)\rangle,
\label{eq:krylov_step}
\end{equation}
where $\exp(-i T_m\Delta t)$ is computed by direct diagonalization of the $m\times m$ tridiagonal matrix $T_m$. We use $m = 60$ and a step $\Delta t \cdot U = 5\times 10^{-2}$, which keeps $\big|\,\|\widetilde\psi(t)\| - 1\,\big| < 3\times 10^{-12}$ across the entire window $tU \in [0,10]$ shown in Fig.~\ref{fig:fig2} of the main text. Smaller steps and larger $m$ produce indistinguishable results.

The Ramsey contrast is finally extracted from the first component of $|\psi(t)\rangle = T^{-1}|\widetilde\psi(t)\rangle$. The curves of Fig.~\ref{fig:fig2} of the main text are converged with respect to the local cutoff and the number of Bogoliubov branches: increasing $N_{\rm loc}$ from $7$ to $9$ or retaining all physical branches changes $|A(t)|$ by less than $10^{-5}$. The finite lattice gives the largest correction. Increasing $L$ from $10$ to $12$ changes $|A(t)|$ by at most $10^{-2}$ and shifts the first minimum and the revival of the QCP curve by at most $0.1/U$, so that the qualitative structure (MI plateau, QCP oscillation, SF collapse-and-revival) is unaffected.
\begin{figure}[t]
\centering
\includegraphics[width=0.95\columnwidth]{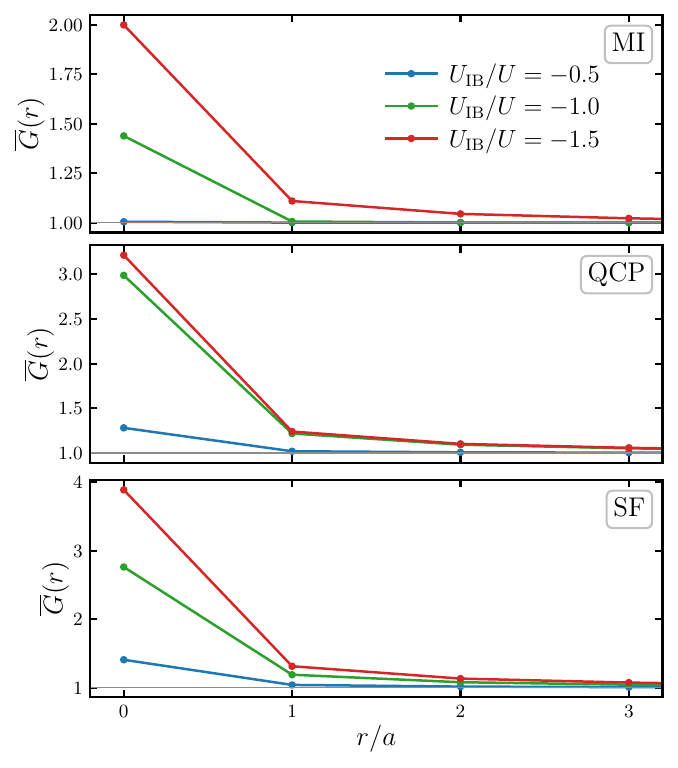}
\caption{\label{fig:fig3-sm}\textbf{Disk-averaged polaron density cloud.} Disk-averaged impurity-conditioned density $\overline G(r)$ over the lattice sites within radius $r$ of the impurity, with the local correlator $G(\br)$ defined in Eq.~\eqref{eq:cloud-def}. Panels show the three host phases, MI at $4J/U=0.05$, near the QCP at $4J/U=0.175$, and SF at $4J/U=0.30$, each at three impurity couplings ($\UIB/U=-0.5$ in blue, $-1.0$ in green, $-1.5$ in red). The thin grey line marks the bulk density. The on-site response evolves from quantized values in the gapped MI to a smooth monotonic enhancement in the SF, while the QCP saturates the response at moderate $|\UIB|$.}
\end{figure}

\section{Polaron density cloud}
\label{sec:cloud}

The bath density operator at lattice site $\br$ has the GW-Bogoliubov representation
\begin{equation}
\op n(\br) \;=\; \bar n + \frac{1}{\sqrt M}\!\sum_s\big[\,N_{s}\,\op b_{s}\,e^{-i\bk_s\cdot\br} + \mathrm{h.c.}\,\big] + \cdots,
\end{equation}
where the second-order terms generate normal one-body and pair contributions to $\delta\op n(\br) = \op n(\br) - \bar n$ when traced against $|\psi\rangle$. Because the variational state carries total momentum $\bs P=\bs 0$, the bare expectation $\langle\psi|\delta\op n(\br)|\psi\rangle$ is translationally invariant and would collapse to a single uniform offset because the impurity contractions $\langle 0|\op c_{\bs 0}\op c^\dagger_{-\bk_s}|0\rangle = \delta_{\bk_s,\bs 0}$, $\langle 0|\op c_{\bk}\op c^\dagger_{\bk'}|0\rangle = \delta_{\bk,\bk'}$ etc.~kill the cross-momentum sums in the AB, BB, AC, BC, CC channels. The physical impurity-resolved cloud is therefore the impurity-bath correlation function
\begin{equation}
G(\br) \;=\; M\,\langle\psi|\,\op n_I(\bs 0)\,\op n(\br)\,|\psi\rangle,
\label{eq:cloud-def-sm}
\end{equation}
with $M\op n_I(\bs 0) = \sum_{\bk,\bk'}\op c^\dagger_{\bk}\op c_{\bk'}$ the localized-impurity projector at $\br_I=\bs 0$. The prefactor $M=1/\langle\op n_I(\bs 0)\rangle$ normalizes by the mean impurity occupation, so $G(\br)$ is the bath density conditioned on the impurity at $\bs 0$ and reduces to the bulk density far away. Equivalently, this is what one obtains by replacing the plane-wave impurity in $|\psi\rangle$ by its localized counterpart $\op c^\dagger_{\br_I=\bs 0}=M^{-1/2}\sum_{\bk}\op c^\dagger_{\bk}$ and evaluating $\langle\op n(\br)\rangle$ in that state. With this definition the operator $M\op n_I(\bs 0)$ replaces the impurity contractions $\delta_{\bk_s,\bs 0}$, $\delta_{\bk_s,\bk_t}$, $\delta_{\bk_s,-\bk_t}$ by unrestricted sums over the bath-mode momenta and reinstates the spatial structure $e^{\pm i\bk_s\cdot\br}$. The induced part of $G(\br)$ then splits into five physically distinct pieces according to which sectors of $|\psi\rangle$ they connect. In terms of the normalized amplitudes of Eq.~\eqref{eq:Hstd},
\begin{align}
G_{AB}(\br) &= \frac{2}{\sqrt{M}}\,\Re\!\left[\widetilde A^{*}\!\sum_s e^{-i\bk_s\cdot\br}\,N_s^{*}\,\widetilde B_s\right],
\label{eq:dnAB}\\
G_{AC}(\br) &= \frac{\sqrt{2}}{M}\,\Re\!\left[\widetilde A^{*}\!\sum_{s,t} e^{-i(\bk_s+\bk_t)\cdot\br}\,W^{\rm sym\,*}_{s,t}\,\widetilde C_{s,t}\right],
\label{eq:dnAC}\\
G_{BB}(\br) &= \frac{1}{M}\,\Re\!\sum_{s,t} e^{-i(\bk_s-\bk_t)\cdot\br}\,K_{s,t}\,\widetilde B_s^{*}\,\widetilde B_t,
\label{eq:dnBB}\\
G_{BC}(\br) &= 2\sqrt{\frac{2}{M}}\,\Re\!\sum_{s,u} e^{-i\bk_u\cdot\br}\,N_u^{*}\,\widetilde B_s^{*}\,\widetilde C_{s,u},
\label{eq:dnBC}\\
G_{CC}(\br) &= \frac{1}{M}\,\Re\!\sum_{s,t} e^{-i(\bk_s-\bk_t)\cdot\br}\,K_{s,t}\,\rho^{C}_{s,t},
\label{eq:dnCC}
\end{align}
with the two-phonon one-body density matrix
\begin{equation}
\rho^{C}_{s,t} \;=\; 2\sum_{u}\,\widetilde C^{*}_{s,u}\,\widetilde C_{t,u},
\label{eq:rhoC}
\end{equation}
\begin{figure}[t]
\centering
\includegraphics[width=0.95\columnwidth]{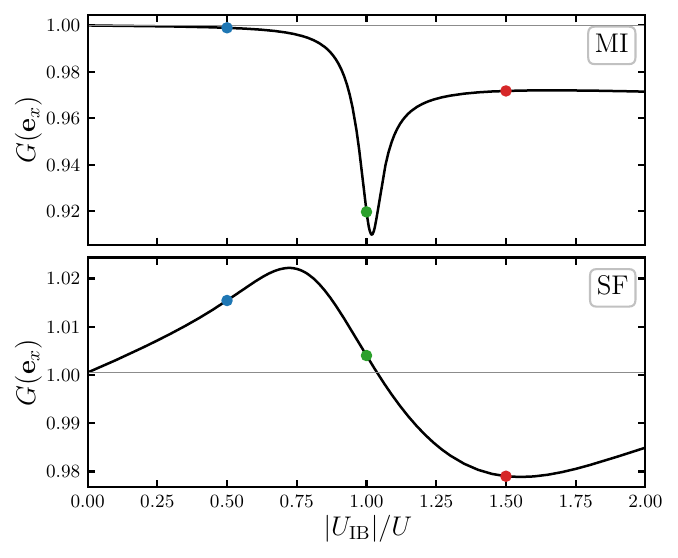}
\caption{\label{fig:g1-sm}\textbf{Nearest-neighbor density.} Impurity-conditioned density $G(\mathbf e_x)$ on a site neighboring the impurity as a function of the impurity-boson attraction, in the MI at $4J/U=0.05$ (top) and in the SF at $4J/U=0.30$ (bottom). The dots mark the couplings $\UIB/U=-0.5$ (blue), $-1.0$ (green), and $-1.5$ (red) shown in Fig.~\ref{fig:fig3} of the main text, and the thin grey line marks the bulk density.}
\end{figure}
which is positive semi-definite and reduces to a one-body density in the Chevy two-phonon sector. The full conditional density is the uniform background plus these induced pieces,
\begin{equation}
G(\br) \;=\; \bar n_{\rm eff} + \sum_{X}\!G_{X}(\br),
\label{eq:dn_total}
\end{equation}
with $X\in\{AB,AC,BB,BC,CC\}$. The momentum sums run over the $L\times L$ grid $k_\alpha=-\pi+2\pi(j+\tfrac12)/L$, $j=0,\dots,L-1$, and $\br$ is the displacement from the impurity with components $-L/2<r_\alpha\le L/2$.
Equations~\eqref{eq:dnAB}--\eqref{eq:rhoC} give the site-resolved maps $G(\br)$ of Fig.~\ref{fig:fig3} of the main text and their disk average in Fig.~\ref{fig:fig3-sm}. Deep in the MI the linear vertex vanishes, $N_s=0$, so that $G_{AB}=G_{BC}=0$, and the one-phonon sector decouples from the ground state, so that $G_{BB}=0$ as well. The MI cloud therefore comes entirely from the pair term $G_{AC}$ and the two-phonon term $G_{CC}$, while in the SF all five pieces contribute.

\paragraph{Atomic-limit occupation staircase.} The on-site response in the MI can be understood from the atomic limit. When the impurity site contains $m$ bath bosons, its interaction energy is
$
E_m=\frac{U}{2}m(m-1)+m\UIB.
$
Writing $g=|\UIB|$, the energy required to add another boson is
$
E_{m+1}-E_m=mU-g.
$
The ground-state occupation is therefore $m=\ell$ for $(\ell-1)U<g<\ell U$, while the states with $m=\ell$ and $m=\ell+1$ are degenerate at $g=\ell U$. The occupation remains finite at every finite $g/U$ because the repulsive contribution grows quadratically with $m$, whereas the attractive contribution grows linearly. The following comparison is made at fixed total bath particle number and unit filling, so the chemical-potential contributions cancel when bosons are transferred from the background to the impurity site. Within the two-phonon ansatz only the first step of this staircase can be described, since transferring one boson from the Mott background to the impurity site requires a particle-hole pair, while each further boson requires an additional pair.

\paragraph{Integrated cloud and ensemble dependence.} Let
$
\hat N_B=\sum_{\br}\hat n(\br)
$
denote the total bath particle number. Translational invariance of the single-impurity state gives
\[
\sum_{\br}G(\br)
=
\frac{
\langle\hat n_I(\bs 0)\hat N_B\rangle
}{
\langle\hat n_I(\bs 0)\rangle
}
=
\langle\hat N_B\rangle_{\rm imp}.
\]
At fixed chemical potential, the integrated cloud is therefore
$
\Delta N_{\rm gc}
=
\sum_{\br}\bigl[G(\br)-\bar n_{\rm eff}\bigr]
=
\langle\hat N_B\rangle_{\rm imp}
-
\langle\hat N_B\rangle_0,
$
where $\langle\hat N_B\rangle_0=M\bar n_{\rm eff}$ is the bath population without the impurity. If $\Omega_{\rm pol}=\Omega_{\rm imp}-\Omega_0$ is the impurity contribution to the zero-temperature grand potential, the thermodynamic identity
$
\langle\hat N_B\rangle
=
-\frac{\partial\Omega}{\partial\mu}
$
gives
$
\Delta N_{\rm gc}
=
-\frac{\partial\Omega_{\rm pol}}{\partial\mu}.
$
The weak-coupling energy shift is
$
\Omega_{\rm pol}
=
\UIB\bar n+\mathcal O(\UIB^2),
$
and hence
\[
\Delta N_{\rm gc}
=
-\UIB\frac{\partial\bar n}{\partial\mu}
+\mathcal O(\UIB^2)
=
-\UIB\kappa+\mathcal O(\UIB^2).
\]
For attractive coupling, $\UIB<0$, this leading contribution is positive whenever the bath is compressible. For $4J/U=0.30$ and $\UIB/U=-10^{-3}$ we find $\Delta N_{\rm gc}/|\UIB|=0.81$, in agreement with $\kappa=0.81$. Deviations at stronger attraction reflect nonlinear dressing and changes of the lowest polaron branch.

In the canonical ensemble, $\hat N_B=N_B$ is fixed and the same definition instead gives
$
\sum_{\br}G(\br)=N_B
$
and
$
\Delta N_{\rm can}
=
\sum_{\br}
\left[
G(\br)-\frac{N_B}{M}
\right]
=
0.
$
Local accumulation is then necessarily compensated by depletion away from the impurity. The canonical and grand-canonical clouds can nevertheless have similar short-distance structures because the compensating canonical response may be distributed over the rest of the system. This similarity does not extend to the complete spatial integral.

\paragraph{Nearest-neighbor density.} Figure~\ref{fig:g1-sm} shows the density on the sites neighboring the impurity as a function of $|\UIB|$. In the MI the neighboring sites are always depleted. The depletion is strongest at $|\UIB|\simeq U$, where the configurations with one and two bosons on the impurity site are degenerate and tunneling resonantly transfers a boson from the neighboring sites onto the impurity. Once the extra boson is bound, $G(\mathbf e_x)$ saturates slightly below one, with the corresponding hole spread over the surrounding sites. In the SF, the neighboring sites are instead enhanced for weak and intermediate attraction, by up to about $2\%$ near $|\UIB|\simeq0.7U$. This is the lattice counterpart of the density enhancement around an attractive impurity in a weakly interacting Bose gas~\cite{penaArdilaImpurity2015,dingPolaronsBipolaronsTwodimensional2023}, and at weak coupling its spatial integral is fixed by the compressibility as discussed above. For $|\UIB|\gtrsim U$ the ground state is dominated by the two-phonon sector and $G(\mathbf e_x)$ drops below one, as the bosons accumulated on the impurity site are drawn from its immediate surroundings. In this strong-coupling regime the truncation of Eq.~\eqref{eq:ansatzS} at two phonons is least controlled.

\end{document}